\documentclass[conference]{IEEEtran}
\IEEEoverridecommandlockouts
\usepackage{url}
\usepackage{cite}
\usepackage{amsmath,amssymb,amsfonts}
\usepackage{algorithmic}
\usepackage{graphicx}
\usepackage{textcomp}
\usepackage{xcolor}
\usepackage{hyperref}
\usepackage{todonotes}
\usepackage{layouts} % TODO: Remove
\makeatletter
\newcommand\thefontsize{The current font size is: \f@size pt}
\makeatother
\def\BibTeX{{\rm B\kern-.05em{\sc i\kern-.025em b}\kern-.08em
    T\kern-.1667em\lower.7ex\hbox{E}\kern-.125emX}}
\begin{document}

\title{Simon's Period Finding on a Quantum Annealer
\thanks{We thank the funding agencies listed in the acknowledgements.}
}

\author{
\IEEEauthorblockN{Reece Robertson}
\IEEEauthorblockA{\textit{Department of Computer Science \&} \\ \textit{Electrical Engineering} \\
\textit{Department of Physics} \\
\textit{Quantum Science Institute} \\
\textit{UMBC} \\
Baltimore, USA \\
reeceroberson@umbc.edu \\
0000-0003-1064-0012
}\\
\IEEEauthorblockN{Krzysztof Domino}
\IEEEauthorblockA{\textit{Institute of Theoretical and} \\
\textit{Applied Informatics} \\
\textit{Polish Academy of Sciences} \\
Gliwice, Poland \\
0000-0001-7386-5441}
\and
\IEEEauthorblockN{Emery Doucet}
\IEEEauthorblockA{\textit{Department of Physics} \\
\textit{Quantum Science Institute} \\
\textit{UMBC} \\
Baltimore, USA \\
0000-0002-2693-8553}\\[3.38em]
\IEEEauthorblockN{Bartłomiej Gardas}
\IEEEauthorblockA{\textit{Institute of Theoretical and} \\
\textit{Applied Informatics} \\
\textit{Polish Academy of Sciences} \\
Gliwice, Poland \\
0000-0002-1454-1591}
\and
\IEEEauthorblockN{Zakaria Mzaouali}
\IEEEauthorblockA{\textit{Institut für Theoretische Physik} \\\textit{Universität Tübingen}\\ Auf der Morgenstelle 14, \\72076 Tübingen, Germany.\\
0000-0003-3948-1318}\\[3.38em]
\IEEEauthorblockN{Sebastian Deffner}
\IEEEauthorblockA{\textit{Department of Physics} \\
\textit{Quantum Science Institute} \\
\textit{UMBC} 
\\
Baltimore, USA \\
\textit{National Quantum Laboratory} \\
College Park, USA \\
0000-0003-0504-6932
}
}
\maketitle

\begin{abstract}
Dating to 1994, Simon's period-finding algorithm is among the earliest and most fragile of quantum algorithms.
The algorithm's fragility arises from the requirement that, to solve an $n$ qubit problem, one must fault-tolerantly sample $O(n)$ linearly independent values from a solution space.
In this paper, we study an adiabatic implementation of Simon's algorithm that requires a constant number of successful samples regardless of problem size.
We implement this algorithm on D-Wave hardware and solve problems with up to 298 qubits.
We compare the runtime of classical algorithms to the D-Wave solution to analyze any potential advantage.
\end{abstract}


\section{Introduction}

Many early quantum algorithms fall into the hidden subgroup class of algorithms, which promise exponential speedup for computational problems \cite{Nielsen:2010}.
For instance, Deutsch's, Simon's, and Shor's algorithms solve hidden subgroup problems \cite{Jozsa:1998}.
Shor's algorithm, in particular, took the world by storm since it solves the factoring problem with an exponential speedup over the best-known classical methods \cite{Shor:1994}.
However, these algorithms strongly assume that one can access many fault-tolerant qubits.
Later research has demonstrated that each hidden subgroup algorithm fails when noise is present on quantum devices \cite{robertson2024simonsalgorithmnisqcloud, Cai:2024, Gupta:2024}.
Simon's algorithm, in particular, fails on gate-based hardware for problems involving over 50 qubits---or 10 qubits on hardware where swap gates are required to implement the algorithm \cite{robertson2024simonsalgorithmnisqcloud}.
In this paper, we analyze the performance of Simon's algorithm in the presence of noise when executed on a quantum annealer rather than a traditional quantum circuit computer.
In so doing, we shift the solution frontier for this problem to approximately 300 qubits on quantum annealers (equivalent to 200 qubit problems on gate-based devices that do not require an ancillary register).
This represents an improvement over the gate-based implementation, however, it is still below the performance of classical approaches.
We achieve this advancement primarily by using the quantum annealer to reduce the information required to solve the problem that must be extracted from the quantum system.

Recall that Simon's algorithm relies on a two-to-one black-box function 
\[
f:\{0,1\}^n\rightarrow\{0,1\}^{n-1},
\]
defined on \( n \)-bit strings, where there exists a nonzero \( s\in\{0,1\}^n \) such that 
\[
f(x)=f(x') \iff x=x'\oplus s,
\]
for all \( x,x'\in\{0,1\}^n \), with \( \oplus \) denoting the bitwise exclusive or (XOR) operator. Here, \( s \) is the hidden period of \( f \), and Simon's problem consists of finding \( s \) \cite{simon}. This algorithm has practical implications in cryptography~\cite{Kaplan_2016, Santoli_2017}. Consider a system that inadvertently employs a function with the above 2-to-1 structure. In such a scenario, Simon's algorithm can efficiently determine the secret shift \( s \), exposing a potential vulnerability. Although secure cryptographic protocols are designed to avoid such weaknesses, the conceptual insights provided by Simon's algorithm are invaluable for guiding the development of quantum-safe encryption methods.

Classically, a single query to the black-box oracle $f$ generally tells us nothing about $s$.
To determine $s$, we must repeatedly query the oracle with different input values until we find two inputs that give us the same output.
Once identified, the XOR of these inputs yields $s$.
The expected number of queries required to find two such inputs grows exponentially with the problem size $n$ \cite{Mermin:2007}.

Quantum mechanically, however, in general, a single evaluation of the oracle reveals one bit of information about $s$.
More specifically, one shot of Simon's algorithm samples a bitstring that is orthogonal to $s$; that is, we obtain some
\begin{equation}
    z\in\left\{z\cdot s = 0 | z\in\{0,1\}^n\right\},
    \label{eq:hiddenSubgroup}
\end{equation}
where $\cdot$ denotes the sum of the bitwise product.
As such, if we obtain ($n-1$) linearly-independent nontrivial samples of \eqref{eq:hiddenSubgroup}, then we can construct a system of equations where the only nontrivial solution is $s$.
In a noiseless setting, our expected number of samples grows linearly with $n$ \cite{Mermin:2007}.

Of course, today's Noisy Intermediate-Scale Quantum (NISQ) computers do not constitute a noiseless setting.
When noise is present in our algorithm, we are not guaranteed that our samples are drawn from ($\ref{eq:hiddenSubgroup}$).
Moreover, if our access is limited to the black-box oracle $f$, we cannot distinguish between the samples that are drawn from \eqref{eq:hiddenSubgroup} and those that are not.
Thus, our classical post-processing step becomes a problem of solving a noisy system of Boolean equations.
A few algorithms exist to solve such systems \cite{alekseev2018solving, alekhnovich2011more}; however, the complexity of this process will likely destroy the quantum advantage for noise levels observed on a current device \cite{domino2024baltimorelightraillinkquantum, koniorczyk2025solvingreschedulingproblemsheterogeneous}.

In this work, we present an adiabatic formulation of Simon’s algorithm executed on two noisy quantum annealers manufactured by D-Wave Systems. By reformulating Simon's algorithm as a Quadratic Unconstrained Binary Optimization (QUBO) problem---a representation equivalent to the Ising model used in quantum annealing \cite{glover_quantum_2022}---we prepare a degenerate ground state that departs from encoding the traditional hidden subgroup \eqref{eq:hiddenSubgroup}. Instead, it encapsulates two specific input values \( z \) and \( z' \) from \( \{0,1\}^n \) satisfying \( f(z) = f(z') \) (and \(z \neq z'\)). This approach enables the annealing process to consistently yield both values, thereby finding the solution to Simon's problem. Additional details on the QUBO setup are provided in Section \ref{methods}, with performance comparisons to conventional techniques presented in Section \ref{results}. An extensive overview on harnessing adiabatic quantum computation for hidden subgroup problems is available in \cite{Albash_2018}. A concrete adiabatic version of Simon’s algorithm is proposed in \cite{Hen_2014}; however, its reliance on a non-Ising Hamiltonian limits its execution on quantum annealers. Moreover, several studies have extended annealing techniques to solve integer factorization \cite{suo_quantum_2020, jiang_quantum_2018, li2017highfidelity, peng_factoring_2019, dridi_prime_2017, PhysRevLett.108.130501}, as well as to address challenges such as the graph isomorphism problem \cite{PhysRevA.89.022342, tamascelli_quantum-walk-inspired_2014} and the minimum distance problem \cite{ismail2024quantum}.

\section{Methods}\label{methods}

We formulate an $n$-qubit input oracle for Simon's problem, where the $i^{th}$ output bit is given by the XOR of input bits $i$ and ($i+1$).
That is,
\begin{equation}
    f(x_1,x_2,...,x_n)_i = x_i \oplus x_{i+1} = o_i.
    \label{eq:oracle}
\end{equation}
This is equivalent to a gate-based implementation of Simon's algorithm where the period of the oracle is the string of $n$ 1s.
Due to the chained structure of this problem, it is representative of the simplest oracles to implement in QUBO form.
Building the QUBO requires the addition of an ancilla for each XOR operation, where the value of the ancilla is given by the AND ($\land$) of the two input bits,
\begin{equation}
    g(x_1,x_2,...,x_n)_i = x_i \land x_{i+1} = a_i.
    \label{eq:constraint}
\end{equation}
As per the D-Wave documentation\footnote{\url{https://docs.dwavesys.com/docs/latest/handbook\_reformulating.html}}, an XOR can be encoded in the following QUBO:
\begin{equation}
\begin{aligned}
    Q(x_1,x_2,o,a) = x_1 + x_2 + o + 4a + 2x_1x_2 \\
    - 2(x_1+x_2)o - 4(x_1+x_2)a + 4oa .
\end{aligned}
\end{equation}
We generalize this to $n$ qubits by taking the XOR of each adjacent pair of input qubits, creating a new output and ancilla qubit with each XOR,
\begin{equation}
    \begin{aligned}
        Q(\textbf{x}, \textbf{o}, \textbf{a}) = \sum_{i=1}^{n-1} x_{i} + x_{i+1} + o_{i} + 4a_{i} + 2x_{i}x_{i+1} \\
        - 2(x_{i} + x_{i+1})o_{i} - 4(x_{i} + x_{i+1})a_{i} + 4o_{i}a_{i}.
    \end{aligned}
    \label{eq:generalQUBO}
\end{equation}
In this equation, $\textbf{x}$ is a length $n$ vector representing the state of the input qubits, with entries $x_i$ for $i\in[1, n]$, and $\textbf{o}$ and $\textbf{a}$ represent the states of the length ($n-1$) output and ancilla registers, respectively.\footnote{Consult the appendix for an illustrative example of equations \eqref{eq:generalQUBO} \& \eqref{eq:penalizedGeneralQUBO}.}

\begin{figure}[bt]
    \centering
    \includegraphics[width=\linewidth]{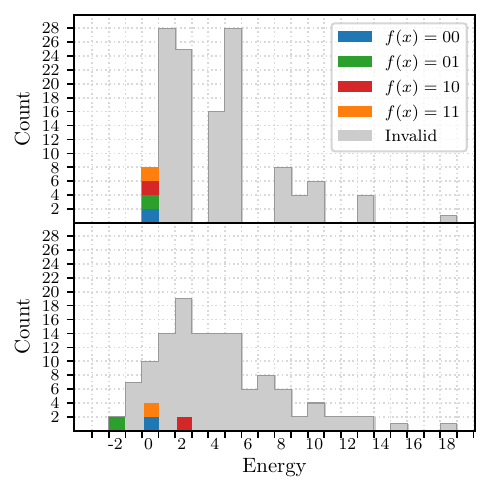}
    \caption{
        Energy spectrum for oracle of size $n=3$ (meaning there are 3 qubits in the input register, and $3n-2=7$ qubits total across all registers).
        The top panel gives the spectrum of the QUBO with no penalties applied.
        In this setting, all the valid evaluations of the oracle appear in a degenerate ground state.
        Hence, sampling this QUBO yields one of these values at random, which queries the oracle with a random input.
        The bottom panel, however, gives the spectrum with penalties $p_1=2$ and $p_2=-2$.
        In this case, two valid oracle evaluations that share a unique output are isolated in the ground state.
        Retrieving both inputs that map to this output is sufficient to solve the problem.
    }
    \label{fig:energySpectrum}
\end{figure}

The spectrum of \eqref{eq:generalQUBO} for the case of $n=3$ is shown in the first panel of Figure \ref{fig:energySpectrum}.
As can be seen, all valid evaluations of the oracle (represented by the colored bars) are grouped in a degenerate ground state.
In this diagram, each non-black color represents a unique output of the oracle.
Notice that these bars have two elements, precisely because the oracle is two-to-one.
In other words, two valid unique assignments of the input, output, and ancilla qubits exist for each oracle output.
Finding both inputs that map to the same output (i.e., sampling both states from one of the colored bars) solves our problem.
However, as we have already observed, in this configuration, all of the valid evaluations of the oracle appear in the same degenerate ground state, regardless of the value of the output qubits.
Therefore, using a quantum annealer to sample from this ground state is equivalent to evaluating the oracle on a randomly chosen input (with the caveat that the input chosen for evaluation is selected miraculously after the computation of the oracle \cite{Mermin:2007}).
Therefore, with \eqref{eq:generalQUBO} alone, we obtain no improvement over the complexity of the classical algorithm.

However, we can do much better with a small addition to the QUBO.
This modification can be made without requiring implementation knowledge of the black-box function.
Consider a modification to \eqref{eq:generalQUBO} where we introduce a penalty parameter on each output qubit:
\begin{equation}
    \begin{aligned}
        Q(\textbf{x}, \textbf{o}, \textbf{a}| \textbf{p}) = \sum_{i=1}^{n-1} x_{i} + x_{i+1} + (1 + p_{i})o_{i} + 4a_{i} + 2x_{i}x_{i+1} \\
        - 2(x_{i} + x_{i+1})o_{i} - 4(x_{i} + x_{i+1})a_{i} + 4o_{i}a_{i}.
    \end{aligned}
    \label{eq:penalizedGeneralQUBO}
\end{equation}
This penalizes state transitions between output states, which effectively fixes a value of the output register throughout our computation.
Significantly, state transitions between the input and ancilla registers occur with the same ease as before, with constraints applied only to enforce the rules of the oracle.
The net effect of this addition is a separation of the degenerate ground state, as shown in the second panel of Figure \ref{fig:energySpectrum} for the $n=3$ problem with $p_1=2$ and $p_2=-2$.
Observe that this modification alters the degenerate ground state---it now contains the valid oracle evaluations for only a single output.
Sampling both states gives us the two inputs that the oracle maps to this output and is a solution to the problem.

The crux of our approach lies in applying penalty parameters to the output transitions within a QUBO formulation of Simon's problem.
This allows us to solve the problem on a quantum annealer by efficiently finding two inputs that the oracle maps to the same output.
We do not need to guess inputs randomly until we find two that match, nor do we need to construct a linear system of equations that we solve to find the period.
We can attain the period directly by sampling both states of a precisely prepared ground state.

\begin{figure}[tb]
    \centering
    \vspace*{0.25cm}
    \includegraphics[width=\linewidth,trim={6cm 2.8cm 6cm 3cm},clip]{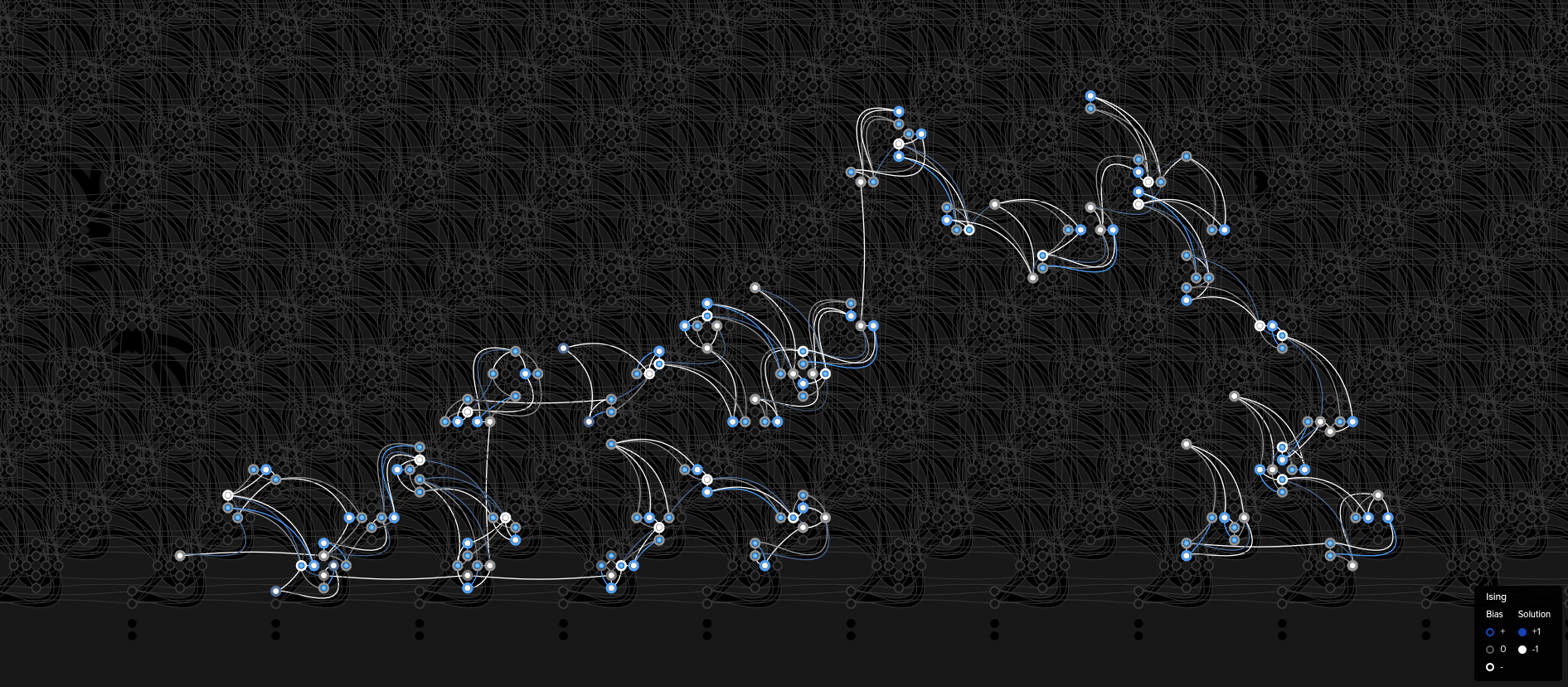}
    \caption{
        The embedding of the QUBO \eqref{eq:penalizedGeneralQUBO} on D-Wave Advantage 5.4 for $n=50$ (50 qubit input to the oracle, 148 total qubits).
        The structure of the QUBO is such that each variable interacts directly with three or six others.
        This connectivity pattern is sufficiently sparse to embed the problem directly on a D-Wave device without introducing swap operations or ancilla qubits.
    }
    \label{fig:embedding}
\end{figure}

We evaluate these QUBOs on the D-Wave Advantage 5.4 (Pegasus) and Advantage2 Prototype 2.6 (Zephyr) quantum hardware.
Each variable in the QUBO interacts with up to six other variables, allowing for a chained embedding within the hardware topology that does not require extra ancillary qubits or swap operations.
See Figure \ref{fig:embedding} for an example embedding of the $n=50$ qubit QUBO into the hardware (and recall that $n=50$ defines the size of the oracle input, and $3n-2=148$ denotes the total number of qubits utilized in the problem).

\section{Results}\label{results}

% \thefontsize
% \printinunitsof{in}\prntlen{\textwidth}
We evaluated the performance of our adiabatic implementation of Simon’s algorithm on the D-Wave Advantage and Advantage2 systems by running problem instances formulated as in \eqref{eq:penalizedGeneralQUBO} for various oracle sizes. Using this QUBO encoding means that the secret string was set to all 1s for all problem sizes. Systematic experiments were conducted by varying both the annealing time and the penalty parameters, with an annealing duration of 100\(\mu\)s proving sufficient to solve the problem reliably over the range of sizes tested.

The proper calibration of the penalty parameters was found to be essential for the algorithm’s success, with penalty values of moderate magnitude proving optimal. When the penalty magnitude was too low (\(\leq1\)) or too high (\(\geq n\)), the solution space degenerated into a near-uniform sampling over all possible outputs, compromising the algorithm’s discriminative ability. The performance of the algorithm is robust for intermediate penalty values; therefore, the remainder of the experiments employed a magnitude of 2.

A further observation is that ground state pairs exhibiting an imbalance in the number of 1s are unevenly sampled. Specifically, states with fewer 1s occur more frequently due to their increased resilience against local noise. To counteract this imbalance, alternating the sign of successive penalty parameters achieved a balanced distribution between the two degenerate ground states. An alternative strategy employing a random sign assignment of penalty parameters yielded comparable performance, and hence, a balanced assignment is reported here as representative of the best-case scenario.

This is illustrated in Table 1, which give the percentage of occurrence of the two target outputs for each configuration of penalty parameters. All experiments were conducted with 4,000 shots on the Advantage2 Prototype 2.6 device. The state $p(z)$ represents the target state beginning with a 0, while the state $p(z')$ represents the target state beginning with a 1. Observe that for the uniform penalty assignment the percentage of occurrence between both states quickly favors $p(z)$, and by $n=50$ the state $p(z')$ is no longer observed.
Under the other assignments the percentage of occurance between both states is more roughly comparable.

\begin{table}[tb]
    \centering
    \begin{tabular}{|c||c|c||c|c||c|c|}
        \hline
         & \multicolumn{2}{|c||}{Balanced} & \multicolumn{2}{|c||}{Random} & \multicolumn{2}{|c|}{Uniform} \\
        \hline
        $n$ & $p(z)$ & $p(z')$ & $p(z)$ & $p(z')$ & $p(z)$ & $p(z')$ \\
        \hline
        \hline
        5 & 39.90\% & 58.25\% & 61.15\% & 35.68\% & 55.48\% & 41.88\% \\
        \hline
        10 & 70.28\% & 18.93\% & 52.10\% & 36.40\% & 62.15\% & 24.88\% \\
        \hline
        15 & 15.95\% & 45.25\% & 24.63\% & 39.68\% & 47.35\% & 6.68\% \\
        \hline
        20 & 14.10\% & 38.05\% & 39.23\% & 5.73\% & 56.95\% & 1.80\% \\
        \hline
        25 & 4.70\% & 30.05\% & 12.83\% & 24.30\% & 36.95\% & 3.05\% \\
        \hline
        30 & 8.55\% & 8.58\% & 8.03\% & 8.05\% & 24.53\% & 3.88\% \\
        \hline
        35 & 1.28\% & 20.38\% & 5.35\% & 6.18\% & 21.83\% & 0.88\% \\
        \hline
        40 & 1.55\% & 7.10\% & 13.85\% & 1.03\% & 9.68\% & 0.63\% \\
        \hline
        45 & 0.50\% & 9.55\% & 8.23\% & 0.23\% & 25.70\% & 0.15\% \\
        \hline
        50 & 4.55\% & 1.30\% & 1.18\% & 1.28\% & 26.55\% & 0\% \\
        \hline
    \end{tabular}
    \vspace{.1cm}
    \caption{Success Percentages by Penalty Configuration on Advantage2}
    \label{tab:penalties}
\end{table}

\begin{figure}[tb]
    \centering
    \includegraphics[width=\linewidth]{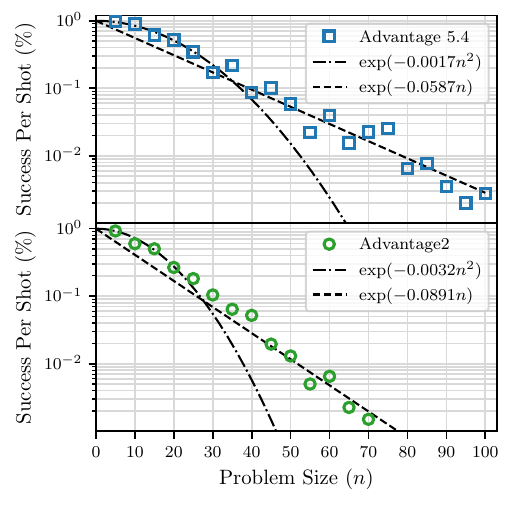}
    \caption{
        Success percentage for our adiabatic Simon's algorithm as a function of problem size.
        As before, the problem size ($n$) denotes the number of qubits in the input register; the total number is $3n-2$.
        The number of shots per problem is fixed at 4000, with an annealing time of 100$\mu s$.
        The experiment was executed on the D-Wave Advantage2 Prototype 2.6 and Advantage 5.4 systems.
        A successful run samples both target values in the degenerate ground state at least once.
        We find that for $n \leq 40$, the success rate is well fit by a Gaussian curve, whereas for $n > 40$, an exponential curve provides a better fit. 
    }
    \label{fig:successRate}
\end{figure}

Results for penalty parameters given by $p_i = 2(-1)^{i+1}$ (the balanced configuration) are given in Figure \ref{fig:successRate}.
To obtain these results, we ran 4,000 shots for a total annealing time of 100$\mu s$  of \eqref{eq:penalizedGeneralQUBO} on a D-Wave Advantage system.
This plot depicts the success rate as a function of problem size, where the success rate is given by the percentage of shots that successfully sampled one of the two values in the degenerate ground state.
(Recall that finding both states with high probability constitutes a solution to the problem.)
We find that the data is well fit by a Gaussian curve for small problem sizes ($n\leq40$), while an exponential curve better approximates the data for larger problem sizes ($40<n\leq100$).
Extrapolating this curve, we expect this function to scale exponentially asymptotically.

\begin{figure}[tb]
    \centering
    \includegraphics[width=\linewidth]{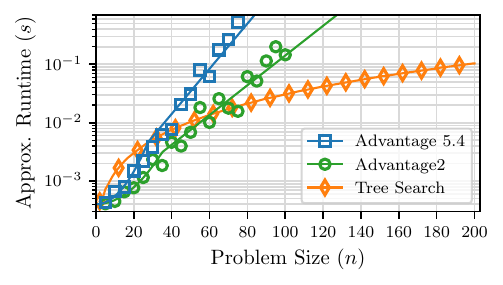}
    \caption{
        Approximate runtime as a function of problem size for our adiabatic Simon's algorithm and a classical tree search QUBO solver implemented by D-Wave.
        The runtime for the annealing algorithm is estimated from the number of shots required to expect to sample from the ground state twice, yielding a $50\%$ chance of solving the problem if the degenerate pair is equally probable.
        The classical algorithm manifests quadratic asymptotic scaling, while our annealing algorithm scales exponentially with problem size.
    }
    \label{fig:runtime}
\end{figure}

In Figure \ref{fig:runtime}, we plot the approximate runtime of our algorithm compared to a classical algorithm.
To estimate the runtime of our algorithm, we computed the expected number of shots required to sample the ground state of the Hamiltonian created by \eqref{eq:penalizedGeneralQUBO} at least twice.
As has been discussed, with correctly selected (randomly assigned) penalty parameters, one can expect that each of the two ground states will be sampled with roughly equal probability.
Therefore, in the best case, sampling the ground state twice yields a $50\%$ chance of sampling both outputs and solving the problem.
If the process fails, one can simply re-run with new random sign assignments for the penalty parameters until success.

For the classical algorithm, we utilized a QUBO solver based on tree decomposition provided by D-Wave as part of the Ocean SDK \cite{DWaveSolverDoc}.
From Figure \ref{fig:runtime}, we can see that the adiabatic version of Simon's algorithm scales exponentially in time, while the classical tree search solver scales quadratically.
The exponential scaling of the adiabatic algorithm arises from the expected number of shots required to sample the ground state at least twice.
The quadratic scaling of the tree search algorithm arises from the tree structure of the search; although the problem space doubles with every new qubit, the classical algorithm needs only to search one additional layer of the tree.

\begin{figure}[tb]
    \centering
    \includegraphics[width=\linewidth]{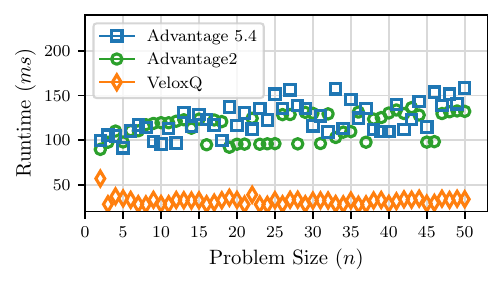}
    \caption{
        Runtime as a function of problem size for our adiabatic Simon's algorithm and the VeloxQ classical QUBO solver, for $n\in[2,50]$ and 1024 shots.
        As before, the experiment was executed on the D-Wave Advantage 5.4 and Advantage2 Prototype 2.6 systems.
        In the region tested, VeloxQ exhibited a slight downward trend in runtime, while the D-Wave systems showed upward trends.
    }
    \label{fig:velox}
\end{figure}

In addition, the performance of our adiabatic approach was benchmarked against VeloxQ, a novel classical solver for QUBO problems that has been shown to outperform other solvers \cite{pawłowski2025veloxqfastefficientqubo}. Figure \ref{fig:velox} compares the runtime for problem sizes \(n\in[2,50]\), using 1024 shots with a 10\(\mu\)s annealing time on the quantum hardware and 1000 optimization steps per shot on VeloxQ. In this range, VeloxQ achieved lower runtimes---with a slight downward trend---and sampled the ground state consistently across all problem instances, whereas the D-Wave quantum annealer began to fail for \(n\geq12\).

Our results show that while the QUBO formulation of Simon’s algorithm is theoretically efficient, the current generation of quantum annealers still faces significant challenges in scaling, as evidenced by the exponential increase in runtime.
Future improvements in quantum hardware, alongside optimized penalty parameter strategies, may be necessary to make adiabatic implementations a competitive alternative to classical methods in solving structured computational problems.

\section{Concluding Remarks}

This paper introduced a new adiabatic version of Simon's algorithm.
Our algorithm prepares a degenerate ground state that encodes two inputs, which the oracle maps to the same output.
Retrieving both of these inputs is sufficient to solve the problem. We implement the algorithm on a D-Wave device for problems with an input register size ($n$) in the range 2--100.
Runtime on the device is quadratic for small problems and exponential for larger problems.
Asymptotically, the runtime of our algorithm scales exponentially. We benchmarked our solution against a classical tree-based QUBO solver implemented by D-Wave and a novel classical QUBO solver called VeloxQ.
The classical algorithms solved the QUBOs quickly, demonstrating a quadratic asymptotic scaling.

Our results indicate that this adiabatic implementation of Simon's algorithm solves the problem in a noisy environment with an advantage over the gate-based circuit method.
In other words, this algorithm is more robust to noise than the traditional implementation.
This performance advantage is realized, in part, by avoiding the expensive post-processing requirement of the gate-based method.
That said, the asymptotic scaling of this adiabatic algorithm is exponential and, therefore, worse than the quadratic scaling of efficient classical algorithms for the problem.
This corroborates prior results on the primacy of classical algorithms in the NISQ era \cite{robertson2024introducinguniqueunconventionalnoiseless}.

This research could be continued in the future.
Most notably, we implemented this algorithm for a Simon's oracle in which the hidden string is the string of all 1s.
One could explore instances of Simon's problem with different hidden strings and evaluate algorithm performance.
This would involve a QUBO of a more complex form than \eqref{eq:generalQUBO}.
We expect that this added complexity would increase runtime for all algorithms comparably, however, it would be worthwhile to verify this conjecture.

\section*{Acknowledgements}\label{acknowledgements}

The authors acknowledge the J\"ulich Supercomputing Centre\footnote{\url{https://www.fz-juelich.de/ias/jsc}} for providing computing time on the D-Wave Advantage™ System JUPSI through the J\"ulich UNified Infrastructure for Quantum computing (JUNIQ).
R.R. acknowledges funding from the UMBC Cybersecurity Institute.
Z.M. acknowledges funding from the National Science Center (NCN) under the Miniatura 8 grant: 2024/08/X/ST2/00099; and from the Ministry of Economic Affairs, Labour and Tourism Baden-Württemberg in the frame of the Competence Center Quantum Computing Baden-Württemberg (project ``KQCBW24").
B.G. acknowledges funding from the National Science Center (NCN), Poland, under
Projects: Sonata Bis 10, No. 2020/38/E/ST3/00269.
S.D. acknowledges support from the John Templeton Foundation under Grant No. 62422.

\appendix

\section{Example QUBO}\label{example}

To aid the understanding of equations \eqref{eq:generalQUBO} and \eqref{eq:penalizedGeneralQUBO}, we present here the derivation of the QUBOs whose energy spectrums are presented in Figure \ref{fig:energySpectrum}.
These QUBOs are given by
\begin{equation}
    \begin{aligned}
        Q(x_1, x_2, x_3, o_1, o_2, a_1, a_2 | p_1, p_2) = \text{ }\text{ }\text{ }\text{ }\text{ }\text{ }\text{ }\text{ }\text{ }\text{ }\text{ }\text{ }\text{ }\text{ }\text{ }\text{ }\text{ }\text{ }\text{ }\text{ }\text{ }\text{ }\text{ }\text{ }\text{ }\\
        x_1 + 2x_2 + x_3 + (1+p_1)o_1 + (1+p_2)o_2 + 4a_1 + 4a_2 \\
        + 2x_1x_2 - 2(x_1 + x_2)o_1 - 4(x_1 + x_2)a_1 + 4o_1a_1 \\
        + 2x_2x_3 - 2(x_2 + x_3)o_2 - 4(x_2 + x_3)a_2 + 4o_2a_2.
    \end{aligned}
\end{equation}
If the first panel of Figure \ref{fig:energySpectrum}, we have that $p_1=p_2=0$, and in the second panel, we have that $p_i=2(-1)^{i+1}$ (i.e., $p_1=2$ \& $p_2=-2$).
It is left to the reader to verify that $Q(0,0,1,0,1,0,0|p_1,p_2)$ and $Q(1,1,0,0,1,1,0|p_1,p_2)$ are in the degenerate ground state under both penalty configurations.

\bibliographystyle{IEEEtran}
\bibliography{bib}

\end{document}